# Privacy Preserving Internet Browsers – Forensic Analysis of Browzar

Christopher Warren[3], Eman El-Sheikh[2] and Nhien-An Le-Khac[1]

[1]School of Computer Science, University College Dublin
Belfield, Dublin 4, Ireland
`an.lekhac@ucd.ie`

[3]RCMP, New Brunswick, Canada
`cewarren15@gmail.com`

[2]University of West Florida, University of West Florida, FL 32514, USA
`eelsheikh@uwf.edu`

**Abstract.** With the advance of technology, Criminal Justice agencies are being confronted with an increased need to investigate crimes perpetuated partially or entirely over the Internet. These types of crime are known as cybercrimes. In order to conceal illegal online activity, criminals often use private browsing features or browsers designed to provide total browsing privacy. The use of private browsing is a common challenge faced in for example child exploitation investigations, which usually originate on the Internet. Although private browsing features are not designed specifically for criminal activity, they have become a valuable tool for criminals looking to conceal their online activity. As such, Technological Crime units often focus their forensic analysis on thoroughly examining the web history on a computer. Private browsing features and browsers often require a more in-depth, post mortem analysis. This often requires the use of multiple tools, as well as different forensic approaches to uncover incriminating evidence. This evidence may be required in a court of law, where analysts are often challenged both on their findings and on the tools and approaches used to recover evidence. However, there are very few research on evaluating of private browsing in terms of privacy preserving as well as forensic acquisition and analysis of privacy preserving internet browsers. Therefore in this chapter, we firstly review the private mode of popular internet browsers. Next, we describe the forensic acquisition and analysis of Browzar, a privacy preserving internet browser and compare it with other popular internet browsers.

# 1    Introduction

Internet security has been a major and increasing concern for many years. Internet security can be compromised not only through the threat of Malware, fraud, system intrusion or damage, but also via the tracking of internet activity. In order to combat these threats, encryption of data as a default setting is now commonplace. Firewalls and Anti-virus programs are essential tools in the fight against computer crime. Indeed, Internet web browsers are daily used by most individuals and can be found on computers, mobile devices, gaming consoles, smart televisions, in wristwatches, cars, and home appliances. Web browsers store user information such as the sites visited, as well as the date and time of internet searches [1]. In addition to clearing their browsing history, users can also prevent storage of such information by using 'private browsing' features and tools [2]. Users may choose to use private browsing features and tools for many reasons, including online gift shopping, testing and debugging web sites, and accessing public computers.

Besides, criminals are using numerous methods to access data in the highly lucrative cybercrime business. Organized crime, as well as individual users, are benefiting from the protection of Virtual Private Networks (VPN) and private browsers, such as Tor, Ice Dragon, Epic Privacy, etc. to carry out illegal activity such as money laundering, drug dealing and the trade of child pornography. Weak security has been identified and exploited in a number of high profile breaches in recent years. Most notably, in 2011, Sony Playstation Network faced a major security breach. Over 77 million Playstation accounts were hacked, resulting in 12 million unencrypted credit card accounts compromised and the site closure for a month. In 2005, the IRS (Internal Revenue Service, USA) faced a data breach that resulted in a reported $50 million in fraudulent claims. As well, in 2015, Ashley Maddison, a site for extra marital affairs, had 37 million account holders' details released. Breaches such as these illuminate the need for better online security and internet privacy.

Additionally, the release of National Security Agency (NSA) and Government Communications Headquarters UK (GCHQ) documentation by Edward Joseph Snowden in 2013 further highlighted the need for improved online security. It is reported that he copied classified information from NSA and Department of Defence (DoD) files containing details of numerous global surveillance programs allegedly run by the NSA and 'Five Eyes' intelligence network with the cooperation of telecommunication companies and European governments. Snowden then released this classified information to the public. The information contained data and emails on foreign targets. Google, Microsoft and Facebook were identified as sources of the information obtained. Following the Snowden breach, there was public outrage at the lack of privacy leading to a rise in the number of browsers offering private browsing. News articles advising on internet privacy assisted in educating the public and a new era of private browsing arose. Although these measures were designed to protect legitimate browsing privacy, they also provided a means to conceal illegal activity.

Although many people use private browsing features for legitimate reasons, some users seek out and exploit third party tools designed for Internet search privacy. Although not marketed solely as private browsers, Mozilla Firefox and Google Chrome have 'private browsing' modes in which the user can browse the Internet without the browser saving Internet histories [3].

On the other hand, there are Internet browsers that are marketed based on its 'total browse privacy'. These browsers claim that for instance they does not save cookies, temporary history files, passwords, or cache. In addition to the privacy claims, these browsers do not require a local installation, and can even be launched from removable media, such as an external hard drive. Their features are very appealing for those seeking complete online anonymity, as well as for those users who intend to commit criminal acts online, such as child exploitation. However, there are very few researches on evaluating of private browsing in terms of privacy preserving as well as forensic acquisition and analysis of privacy preserving feature of "total browse privacy" internet browsers. Therefore, in this chapter, we firstly review the private mode of popular internet browsers. Next, we describe the forensic acquisition and analysis of a privacy preserving internet browsers Browzar, and compare it with popular internet browsers. The rest of this chapter is organised as follows: Section 2 shows the background of this research including related work in this domain. We present the forensic acquisition and analysis methods of Browzar privacy browser in Section 3. We describe our experimental results and discussion of analysing Browzar privacy browser in Section 4. Finally, we conclude and discuss on future work in Section 5.

## 2      Background

The availability and popularity of the "Internet" has grown exponentially since the 1990s. In 1995, an estimated 16 million users had access to the Internet - a mere 0.4% of the world's population. As of December 2015, it was estimated that over 3 billion users had access to the Internet, accounting for approximately 46% of the world's population [4]. People connect to the Internet using multiple devices, multiple times a day. Whether it's on a personal or work computer, a mobile device, or on a smart watch, millions of users perform various online activities every day. One of the most common activities is using the Internet to perform searches for information. The most popular search engine, Google, is estimated to process over 40,000 search queries every second. This translates to over 3.5 billion searches per day, and 1.2 trillion searches per year worldwide [5]. Unfortunately, with its rise in popularity, the Internet has created a new form of criminal activity, often referred to as 'cybercrime'.

Cybercrimes have been defined in various ways, but are generally categorized into one of the following two types of Internet-related crime: 'advanced cybercrime' and 'cyber-enabled crime' [6]. Advanced cybercrimes are attacks against computer hardware and software, whereas cyber-enabled crimes are 'traditional' crimes that have been enabled by and have taken new form with the advent of the Internet. Cyber-enabled crimes include crimes against children, financial crimes, and terrorism. International agencies have identified cybercrimes as an issue. The European Union's law enforcement agency, Europol, for example, notes that the Internet is a major source of criminal activity. Europol suggests that the Internet enables organized crime groups to access large pools of victims and to carry out a diverse range of the criminal acts on large scales and in shorter periods of time [7]. The obscure nature of the criminal activity makes it very challenging for law enforcement not only to catch criminals using the Internet, but also to successfully pursue prosecution. In an effort to avoid detection and prosecution by law enforcement agencies, criminals resort to using various tools and tactics to conceal their illegal online activity. These tools and tactics often leave little to no evidence behind.

### 2.1 Internet Security

Online transactions have become a normal part of everyday internet activity. Whether it is buying, banking or socializing using the desktop, smartphone or tablet, our personal information, documents and habits would be vulnerable to public exposure without internet security. Firewalls and Anti-Virus programs are now essential tools in the fight against cybercrime. As criminals become more aware of and adept at using anti-forensic tools such as encryption, obfuscation and programs such as CCleaner [8] which is used to wipe any trace of data on a hard drive, forensic investigators have to keep up with the ever changing cybercrime landscape. In the past, investigators on warrant were routinely told to 'pull the plug' on any devices found on scene and deliver them to forensic examiners for analysis. To do this now could mean the loss of valuable live data and automatic encryption of the hard drive. Remote wipe applications are now also included on many mobile devices so the judgement call of balancing the need to secure the network to prevent remote wipe codes being transmitted, with the possibility of acquiring data from either a protected drive or cloud based data, has never been more crucial.

The publication, Understanding Internet Security [9] by BigPlanet, states that over 90% of internet users have spyware lurking on their computers without their knowledge, delivered by a number of different methods. Websites carrying malicious code account for a large majority of attacks in the form of adware or spyware. Email attachments can also execute code to infect a user's computer by way of a virus, worm or allow remote access to be established.

### 2.2 Forensic Analysis and Web Browser

According to Oh et al., a crucial component of a digital forensic investigation is searching for evidence left by web browsing. Each movement taken by a suspect using a web browser can leave a recoverable trace on the computer. Analysing this information can produce a variety of artifacts, such as web sites visited, time and frequency of access, and keyword searches performed by the suspect [10].

Oh et al. further state that in a Web browser forensic investigation, the simple parsing of information is not enough. Additional forensic techniques may be required to extract more significant information such as "the keyword searches and the overall user activity". For example, depending on the possible user activities during a single browser session (i.e., email, online banking, blogging), investigators must analyse the information generated from each browser using the same timeline.

Web browser history and analysis has become an increasingly important area of computer forensics. As techniques advance and new tools become available to examine web browsers, investigators are able to reconstruct timelines, identify suspect browsing habits, and recover evidence of crimes. Furthermore, the results uncovered during web browser forensics can determine the objective, methods, and criminal activities of a suspect.

### 2.3  Privacy Browsing

Although private browsing has legitimate uses, such as activity on multiple user devices and political restrictions, many individuals are using the shield of anonymity to carry out illegal activity on the Internet. Private browsing is designed in some web browsers to disable browsing history and the web cache. This allows a user to browse the Web without storing data on their system that could be retrieved by investigators. Privacy mode also disables the storage of data in cookies and browsing history databases. This protection is only available to the local device as it is still possible to identify websites visited by associating the IP (Internet Protocol) address at the website.

Apple introduced the first Internet privacy features in a browser in 2005 in the web browser Safari [11]. Since this time, 'privacy features' or 'privacy modes' have become standard in most Internet web browsers. Google Chrome, Mozilla Firefox and Internet Browser each also have private browsing features.

Boneh, et al. [12] examined private browsing features introduced by four popular browsers: Internet Explorer, Mozilla Firefox, Google Chrome and Apple Safari. The authors noted that private browsing modes have two goals:
  1) to ensure sites visited while browsing in private leave no trace on the user's computer;
  2) to hide a user's identity from web sites they visit by, for example, making it difficult for web sites to link the user's activities in private mode to the user's activities in public mode.

The research also identified the inconsistencies when using private mode with the popular browsers and revealed that, although all major browsers support private browsing, inconsistency in the type of privacy provided by each differs greatly. Firefox and Chrome attempt to protect against both web and local attacks while Safari only prevents local issues [12].

Although legitimate reasons do exist for using private browsing, some users take advantage of these features to conceal criminal activity.

Erik Larkin explains that a typical private browsing session is initiated from within the interface of the main browser [2]. Once initiated, it is carried out in its own private browsing session window until the session is terminated. While in a private browsing session, no updates are made to the browser history, and upon terminating the session, the browser will delete any cookies stored during the session. This will also clear the download list. Larkin suggests that while these features do add privacy, they will not completely cover a user's tracks.

Hedberg's research led to similar findings and suggests that there are many misconceptions about private browsing [13]. To the end user, private modes and private browsers typically perform as advertised. However, using a series of web browser forensic techniques and tools, Hedberg proved that claims of total browse privacy are false. During his research involving the private modes for the Google Chrome, Mozilla Firefox and Microsoft Internet Explorer browsers, Hedberg successfully recovered artifacts for each browser from within the hard drive and memory of the system.

Well known browsers such as Google Chrome, Internet Explorer, Safari and Mozilla Firefox rely on similar methods to ensure speed and popularity of their product. Web Cache is a popular way of storing data that can be easily and quickly accessed, thereby negating the necessity to find data that has already been used. History databases, thumbnails (small stored images), temporary files and cookies (user and site specific data) all help to speed up the user experience and, in their path, leave a plethora of artefact evidence for examiners to feast on. Many studies have been carried out in this area and free tools, such as ChromeHistoryView, ChromeCacheView, IECacheView, as well as forensic software such as Internet Evidence Finder, are available to automate the examination process. All the above browsers have the option to operate in private mode. Research by Khanikekar [14] indicates that the use of Internet Explorer in 'Protected Mode' runs a 'Low Privilege' process, preventing the application writing to areas of the system that require higher privilege.

## 2.4   Browzar

Browzar Internet browser [16] was designed to offer total browse privacy for users. It was designed and marketed for individuals to browse the Internet in a private way – either on a personal or shared computer. Browzar claims that it does not save any information to the local system – and even includes an automatic purging feature to remove any information left behind upon terminating a browsing session. Browzar is an Internet Explorer (IE) shell browser, so Internet Explorer is required on the host system for it to function. Further, it does not require a standard installation; instead, it can be launched from a removable media device, or even from the vendor website. Working in computer forensics, it is common for investigators to rely on the recovery of Internet evidence to solidify investigations and prosecution. Presently, there is little documented research on Browzar and its privacy claims. This chapter will challenge the key selling points advertised by Browzar. Various experiments will be conducted as part of a full forensic analysis to determine if any artifacts are left behind on a system after using Browzar. Additional comparisons will be made to the Google Chrome and Mozilla Firefox browsers to identify similarities in artifact preservation. More details can be found in the following sections.

# 3 Forensic acquisition and analysis of Browzar privacy browser

In this chapter, we aim to answer the following research questions:

1. Is Browzar capable of providing complete privacy to users, avoiding all forensic techniques? With claims of total browse privacy, it is important to confirm whether Browzar performs as advertised simply from an end user's stand point, or if it actually can avoid all forensic techniques, including memory analysis, tools specific to web browser forensics, and a full forensic analysis using software such as X-Ways.
2. How does Browzar privacy compare to Google Chrome and Mozilla Firefox privacy modes? This will help determine if the Browzar product works better or worse than most private modes that are now standard in many Internet browsers. This comparison will also help to identify whether or not Browzar could be deemed an anti-forensic web browser.
3. What evidence was recoverable during post-mortem analysis of Browzar? The most common type of analysis occurs after a device is seized. The post mortem analysis of Browzar will benefit investigators and offer recommendations on analysing data using specific tools such as web browser forensic tools. Further, post mortem analysis enables access to database files and information found in unallocated space – information that would otherwise not be available during a live or logical preview.
4. What evidence was recoverable during live state analysis of Browzar? Investigators today are increasingly facing situations that no longer allow them to unplug a system and perform a traditional hard drive acquisition. The information captured in the memory of a system is considered volatile information - which is information that is lost once the host system is powered off. This data is extremely important and useful as it can often help determine the most recent activities on a computer – including browser specific findings such keyword searches and website history.

## 3.1 Adopted Approach

The following approach was adopted to address the questions posed above. Based on the nature of this research subject, various approaches and tools were used to recover and analyse information.

Change monitoring is a useful technique used to determine the impact a specific program or action has on a system. With respect to web browsers, change monitoring can help determine any modifications that the browser is making to the system, and, in turn, identify the system files and registry keys being changed as a result of a particular action. Further to this, a virtual environment is ideal for testing different versions of a browser and for validating findings.

Live data forensics is an increasingly important aspect of computer forensic investigations. The objective of live data forensics is to minimize impacts to the integrity of data while collecting volatile evidence from the suspect system. The primary source of volatile information is contained within the RAM of a system. Information residing is RAM is stored there on a temporary basis. Once the host system is powered off, the data is cleared from the RAM and lost forever. Forensic tools have been developed that allow investigators to capture and process the data stored in RAM. As part of the experiments performed in this research, the host system will be treated as a real world exhibit. Using forensic techniques and tools, the RAM will be captured in a forensic manner and preserved for further analysis. Various tools will be used to process and analyse the contents of the RAM for web browser specific evidence.

Post mortem analysis often provides the best opportunity to gather the most evidence in a digital forensic investigation. The post mortem forensic process begins with a data acquisition. This involves creating bit by bit forensic duplicates of any digital evidence. These forensic duplicates not only protect the integrity of the original data, but also enable additional information to be recovered from deleted and slack space. Software tools, such as X-Ways, are then used to process and categorize the information as part of the forensic analysis. With respect to web browsers, full forensic software like X-Ways include features designed to parse out web browser databases and any other relevant metadata.

### 3.2 Experimental environments

The results outlined in this minor thesis were obtained using two separate working environments. The first environment was created using an Apple Mac Mini computer. Apples Boot Camp program was used to configure the system to allow a dual boot configuration – thus enabling an installation of the Windows 7 operating system (service pack 1). All available Windows 7 updates – both security and software – were performed using the Windows Update feature to ensure a complete up to date working environment. The operating system was installed on a 75 GB VMware virtual machine (VMware Workstation 12 Pro). The virtual machine served two purposes: to create an isolated working environment and to allow the system to be reverted back to an earlier date (i.e., snapshot) if necessary.

Change monitoring tools including Process Monitor (Procmon) and Index.dat Analyzer were installed in the virtual environment. Microsoft explains Process Monitor as an "advanced monitoring tool for Windows that shows real-time file system, Registry and process/thread activity". Internet Explorer, which is native to Windows 7, also remained as part of the environment. It should be noted that Browzar was tested using Internet Explorer versions 9 through 11. Although compatible with each version, there are instances where results were gathered using only Internet Explorer version 9. Depending on the stability and compatibility, certain results may have been gathered using Internet Explorer versions 10 or 11.

To further analyse the behaviour of each browser, a second testing environment was created using a Dell Latitude E6430 laptop containing a 120 GB OCZ solid state hard drive and 4 GB of RAM. The system was left connected to the Internet for several days, all while actively using various search engines to perform specific web activity. These secure sites required a successful login with a username and password.

After approximately 48 hours, the system memory (RAM) was acquired using the dumpit.exe tool, and the system was successfully shut down using a proper shutdown sequence.

### 3.3 Keyword Search Terms

Figure 1 shows the key search terms and websites used to perform a series of specific searches during the forensic analysis.

| Type | Information |
|---|---|
| URL | ucd.ie |
| URL | espn.com |
| URL | nfl.com |
| URL | paypal.com |
| Search Engine | google.ca |
| Social Media | facebook.com |
| Search Terms (Chrome) | Hockey, Baseball, Golf, Football |
| Search Terms (Firefox) | Brazil, South Africa, Canada |
| Search Terms (Browzar) | Beer, Wine, Scotch |

**Fig.1** Search criteria used during testing

## 4 Experiments and analysis

### 4.1 Change Monitoring

*4.1.1 Browzar*

Contrary to the successful research findings discovered for Google Chrome and Mozilla Firefox, limited information pertaining to Browzar was found. Therefore, the best source of information would be from testing the product itself. Upon completing the initial setup of the virtual machine, Browzar version 2.0 (Windows Style Theme) was downloaded directly from the vendor website. The installation file, as advertised, was around 200kb in size and took only seconds to download in its entirety.

Using Procmon, the file system and registry changes were examined to determine the footprint of the Browzar installation. A Browzar session was initiated in order to visit various websites and to perform the keyword searches outlined above. This activity was designed to populate the system with random Internet artifacts, such as pictures, keyword searches, cookies, etc. Procmon values were created for browzar-winstyle2000.exe and iexplore.exe to help filter results. Although no results were registered for the iexplore.exe value during the launch, plenty of activity was logged for browzarwinstyle2000.exe (Figure 2)

**Fig.2** Activity logs of Browzar

The Browzar interface, originally modelled after Mozilla Firefox, is similar in appearance to a typical Internet browser. Upon further inspection, differences emerged. Various privacy features such as a separate field for private searches, as well as Secure Delete and Force Cleanup options were present (Figure 3).

Similar to other Microsoft Internet Explorer shell browsers, much of Browzar's activity involved the registry, specifically with the following keys:

- HKCU\Software\Microsoft\InternetExplorer
- HKCU\Software\Microsoft\Windows\CurrentVersion\InternetSettings
- HKCU\Software\Wow6432Node\Microsoft\InternetExplorer

A review of the file system activity in Procmon revealed that various files were being created on a temporary basis upon starting Browzar. To illustrate this further, a Browzar folder was created in C:\Users\21\AppData\Roaming which contained a single file called recovery.lck (see Figure 4).

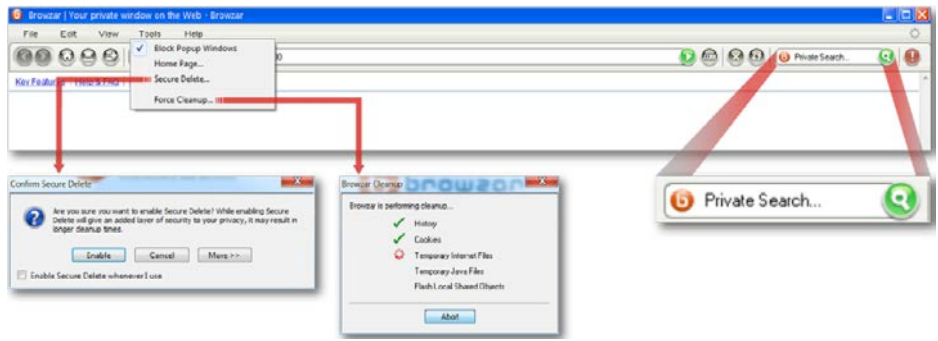

**Fig.3** Browzar privacy features

**Fig.4** Procmon results for BrowzarWinstyle2000.exe

Browzar claims that the recovery.lck contains a date and time stamp of the precise moment the Browzar session began [11]. It states further that this file contains no information about websites visited during a Browzar session. As advertised, any remnants of these findings were purged upon closing the browsing session (Figure 5).

File operations were also monitored using Process Monitor to determine what information was being written to the disk. The majority of information being written during the Browzar session was to the following locations:

- C:\Users\21\AppData\Roaming\Microsoft\Windows\Cookies
- C:\Users\21\AppData\Local\Microsoft\Windows\TemporaryInternetFiles\Content.IE5

The index.dat file, which is located in the Content.IE5 folder, is used by Internet Explorer version 9 as a repository for information such as websites, search queries, and recently opened files (Figure 6).

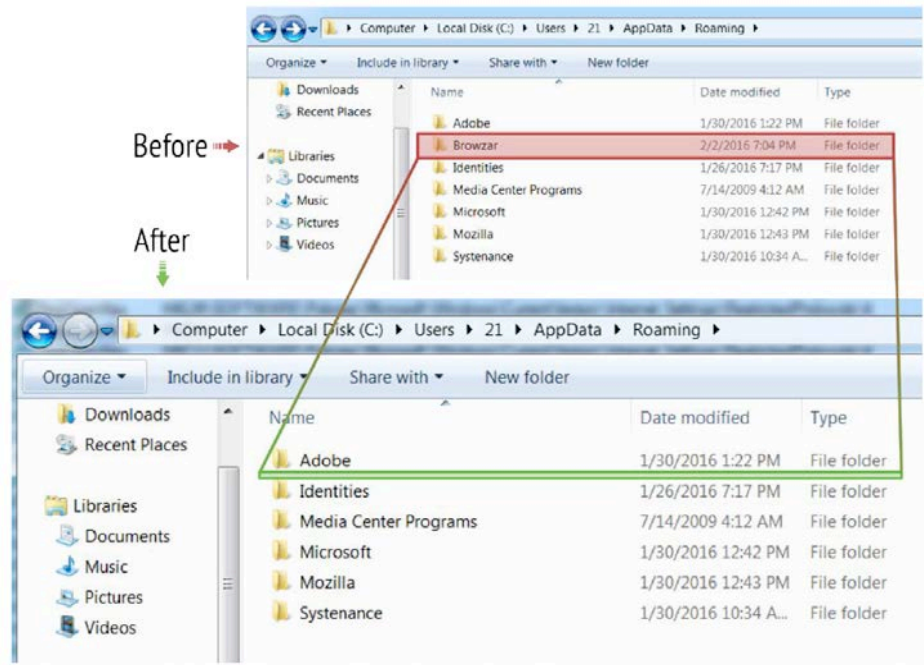

**Fig.5** Before and after illustration of deleted temporary Browzar folder

**Fig.6** Procmon activity showing create, read, write and close operations

During the Browzar session, the index.dat file showed continuous create, read, and write operation activity (Figure 5). Upon terminating the session, additional close operation activity to the index.dat file was observed.

To illustrate this further, the www.nfl.com website was accessed using Browzar. During the browsing session, the index.dat file was accessed using Index.dat Analyzer v2.5. It contained evidence of the www.nfl.com url as well as additional artifacts such as .png and .jpg files from the website. (Figure 7)

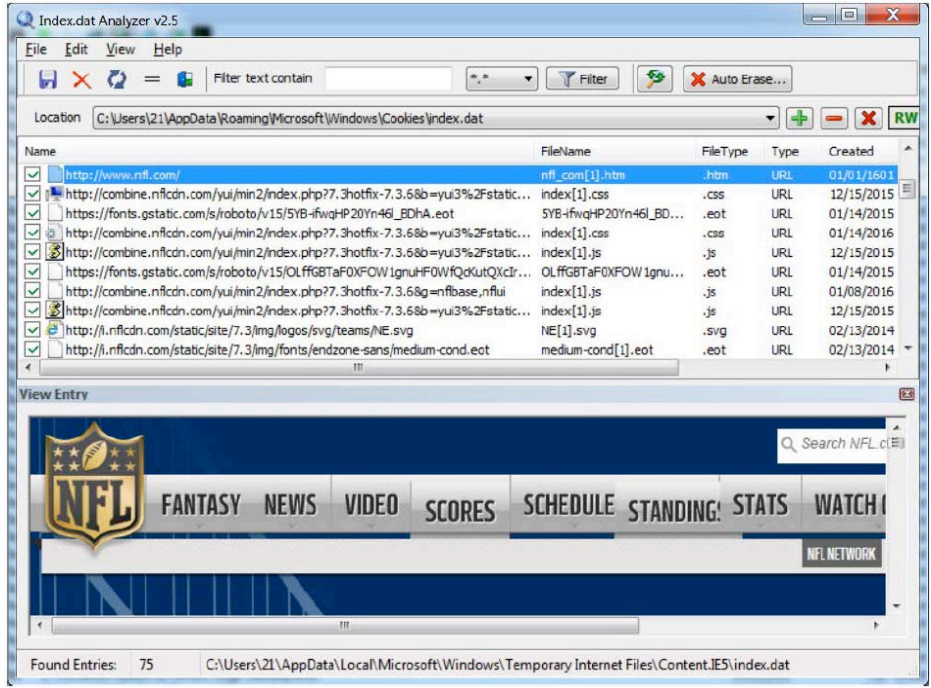

**Fig.7** Index.dat Analyzer results showing www.nfl.com artifacts

Upon closing the browsing session, all the information was removed from the index.dat (Figure 8).

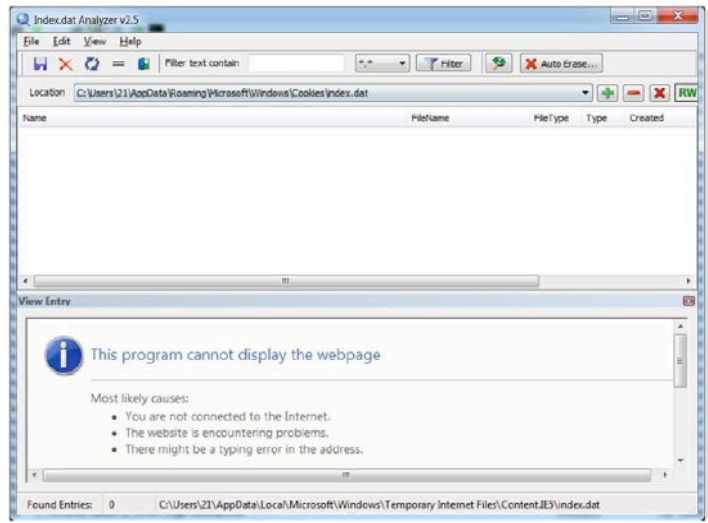

**Fig.8** Index.dat Analyzer showing *www.nfl.com* artifacts were removed

Prior to terminating the Browzar session, the Secure Delete and Force Cleanup options were initiated and then terminated by exiting the browsing session using the File menu.

*4.1.2 Chrome and Firefox*

Using the Chrome and Firefox private modes, the remaining searches and web activity assigned to each browser based on Figure 3 was carried out. Chrome and Firefox browsers were creating and using SQLite databases to store browsing information, as evident from the Procmon results. These SQLite files were analyzed using X-Ways and Oxygen forensic software programs. The Oxygen Forensic software is a mobile forensic software typically used for extracting and analyzing cell phones, smartphones, and other mobile devices. However, it also contains a built in SQLite viewer that parses out and presents the information stored in an SQLite database file. While observing the installation of Chrome and Firefox, it was also documented that user profiles for each browser were generated as part of the installation (see Figure 15). Both browsers allowed for multiple user profiles using their own SQLite database files, enabling individual users the ability to collect their own bookmarks, store passwords, and history.

## 4.2      Forensic acquisition

*4.2.1 Live Data Forensics*

To successfully capture the RAM, a 32 GB Lexar USB thumb drive was forensically wiped and loaded with the dumpit.exe utility. Using the dumpit.exe utility, the RAM capture was initiated, storing the output directly to the previously forensically wiped 32 Lexar USB thumb drive.

*4.2.2 Post-Mortem Data Acquisition*

In order to acquire information from the hard drive of the Dell Laptop, a forensic data acquisition was performed. The hard drive was removed from the laptop and connected via a USB tableau write blocker to a forensic analysis machine. Using FTK Imager version 3.4.2.2 forensic software, the data acquisition process was performed. The electronic information contained on computer storage media must be acquired by making a complete physical copy of every bit of data located on the computer media in a manner that does not alter that information. The integrity of the data is authenticated at the end of the acquisition using the MD5 hashing algorithm. This hashing process creates a 128 bit hash value of the data - which is often referred to in computer forensics as the digital fingerprint - to ensure the acquired electronic information is an exact duplicate of the storage media. For this particular acquisition, the forensic duplicate of the hard drive was created using the Encase image file format - known as E01.

## 4.3 Forensic analysis

*4.3.1 Browzar - Memory Analysis*

During the data processing stage, IEF was used to process the RAM and search for artifacts pertaining to the Browzar specific search strings outlined in Figure 3. IEF recovered several Google Searches performed using Browzar (Figure 9). By further analysing the memory using X-Ways, additional keyword searches were identified. For example, through X-Ways, six instances of the term "beer" were recovered.

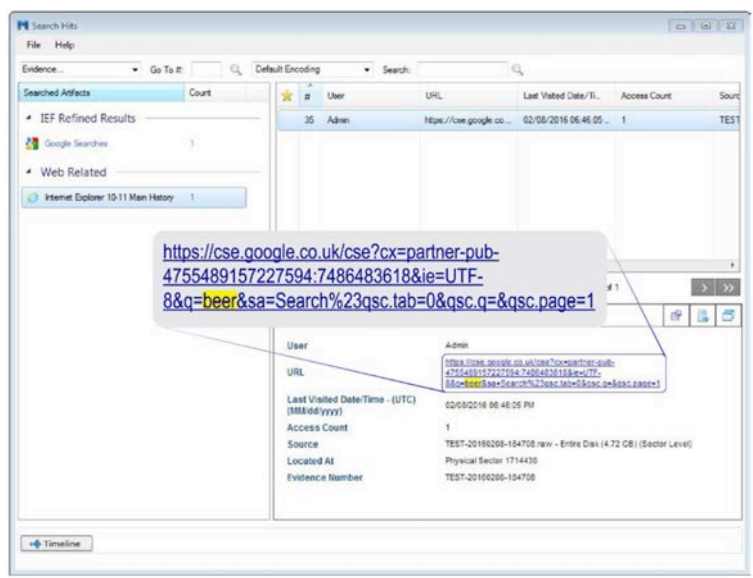

**Fig.9** IEF results showing recovered google search for "beer"

Similar results were also identified for the other search terms. Residual artifacts found in RAM such as these are often left behind while the browsing session remains active and stored in RAM. By performing this additional examination, it illustrates the importance for forensic analysts to perform manual data checks and also the importance of validating findings using a second forensic tool.

*4.3.2 Browzar – Post Mortem Analysis*

Considerable findings were also uncovered during post mortem analysis for Browzar in the pagefile.sys. The pagefile is used by the Windows operating system as an extension of random access memory. For example, the pagefile is often accessed when a computer system's RAM is completely full. The operating system views this file as actual physical memory, allowing for quicker access to applications stored within it compared to if they remained in their original location. In this instance, Pagefile findings were recovered using IEF and X-Ways and included keyword searches, web history, and images from webpages.

In addition to the findings in the pagefile, several instances of the Browzar folder were discovered (see Figure 6). One of the instances containing a recover.lck file was examined. The majority of this file was not legible in X-Ways and it did not contain any stored websites - as claimed by Browzar [10]. However, one section that referenced a WebCachev01.tmp file was discovered (Figure 10).

**Fig.10** WebCachev01.tmp reference found in recovery.lck file

This prompted the researcher to locate and analyze the WebCachev01.dat file. The WebCachev01.dat file is used by Internet Explorer (version 10 and higher) to maintain the web cache, history, and cookies. A review of this file in X-Ways uncovered more references to the keywords used during the Browzar session, confirming that Browzar is leaving traces of activity behind (Figure 11). In addition to these findings, Browzar's private search feature revealed that all searches were performed using a *www.google.co.uk* address.

**Fig.11** Evidence of Browzar searches found in WebCachev01.dat

To investigate this further, Wireshark tool was used to perform a series of tests. Wireshark is a free and open source network protocol analyzer for Unix and Windows. By capturing the browsing session using Wireshark, the data packets sent to and from the test environment during the Browzar session were reviewed and filtered. Based on the results of these tests, it was evident that Browzar was not leaking any session information and that the private search feature found on the Browzar interface was simply a "Google Custom Search" engine – a feature that was hardcoded into the tool.

*4.3.3 Google Chrome - Memory Analysis*

Although minimal, several findings of interest relating to the Chrome private session were recovered during the analysis of the system memory. By filtering the results using the keyword "football", IEF recovered a user generated Google search involving this word. Similar discoveries were made during a manual examination using X-Ways (Figure 12).

**Fig.12** Sample Chrome results found using X-Ways

*4.3.4 Google Chrome – Post Mortem Analysis*

The Pagefile.sys file contained some of the most valuable information recovered from the Chrome private browsing session. For example, various keyword searches used during the Chrome session were recovered in the pagefile. In addition to keyword searches, urls, bookmarks, and images were all discovered in the pagefile. To illustrate this further, the evidence returned by IEF was filtered for the search term "baseball" and it hit on the following 10 images. Several bookmarks were created during the private browsing session. It is to be noted that these bookmarks remained available during subsequent normal (not private) browsing sessions. These were easily recoverable using IEF and X-Ways via bookmarks SQLite database file (Figure 13).

**Fig.13** Illustration showing references to Chrome bookmarks in X-Ways and in Chrome

The remaining Chrome SQLite databases were examined, but did not contain any information related to the private browsing session. Aside from the residual content left behind in memory and the spillage found in the pagefile, Chrome did a good job of discarding any traces of activity, and certainly performed as advertised from the viewpoint of the end user.

*4.3.5 Mozilla Firefox - Memory Analysis*

IEF was used to process the RAM and search for evidence pertaining to the Firefox specific criteria outlined in Figure 3. IEF was able to recover several Yahoo searches performed during the Firefox browsing session. In addition to keyword searches, urls, bookmarks and images were also identified among the IEF results.

*4.3.5 Mozilla Firefox – Post Mortem Analysis*

Post mortem analysis was performed for Firefox using X-Ways. Once again, the pagefile.sys contained a wealth of information. From within the pagefile, urls, parsed search queries, and pictures were recovered (Figure 14). To gain a fuller picture of the activity performed during the Firefox session, evidence of a search for "South Africa" extended beyond the recovered parsed search queries. For example, multiple images – both full and partial - were recovered from the pagefile that originated while browsing South African travel websites. With respect to bookmarks, several were created during the private browsing session. It is to be noted that these bookmarks remained available during subsequent normal (not private) browsing sessions (Figure 15).

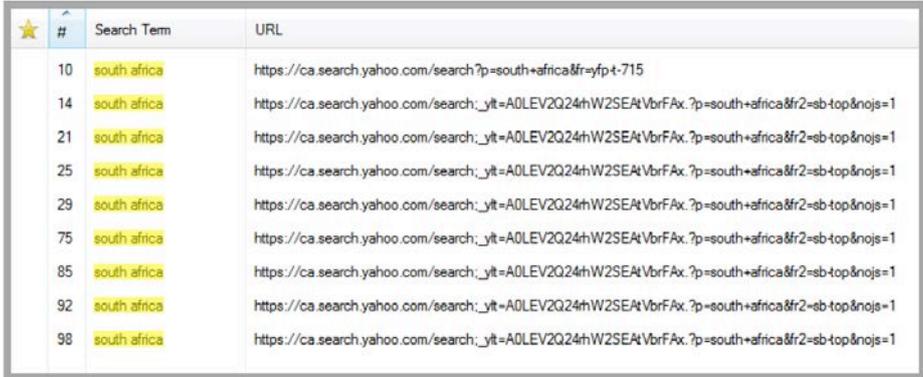

**Fig.14** Sample IEF result showing searches for South Africa

The remaining Firefox SQLite databases were examined, but did not contain any information related to the private browsing session. Aside from the residual content left behind in memory, and the information found in the pagefile, Firefox did not store information on the system. From an end user's perspective, the Firefox private browsing feature also performed as advertised.

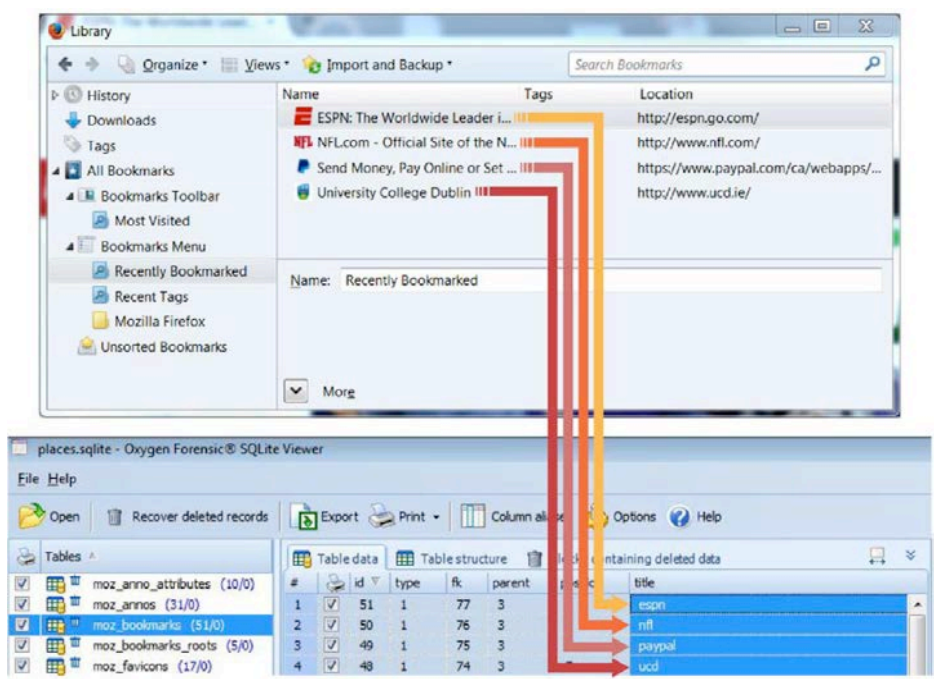

**Fig.15** Illustration showing references to Firefox bookmarks in Firefox and Oxygen SQLite Viewer

### 4.4 Discussion

The results of this chapter will serve to fill the gap in knowledge regarding the forensic analysis of the browser, Browzar. Based on the results, it was concluded that private browsing artifacts are recoverable. All three browsers tested as part of this minor thesis reduced the amount of information left behind after usage, and were sufficient at minimizing and preventing the amount of data stored on the host system. However, none of the browsers tested were capable of fully preventing or controlling what information remained in the memory of the system and the pagefile. In this section, we discuss on the research questions we raised in Section 3:

1. Is Browzar capable of providing complete privacy to users, avoiding all forensic techniques?

    To the end user, the product performed as advertised. Upon closing the browsing session, it removed all traces of web browser activity. However, using a combination of forensic tools and techniques, evidence, including pictures, keyword searches, and urls, were easily recovered in both the memory and in the pagefile of the test system.

2. How does Browzar privacy compare to Google Chrome and Mozilla Firefox privacy modes?

    Out of the three browsers forensically examined for this minor thesis, Browzar was found to leave the most information behind. Not only were artifacts recovered from various locations, it also left a folder named Browzar behind on the system. This folder was instrumental in pointing to further evidence, leading to a more successful forensic examination.

3. What evidence was recoverable during live state analysis of Browzar?
   Live analysis, although not always an option available to investigators proved to contain valuable evidence in this case. From within RAM, evidence of keyword searching, websites visited, and pictures were recovered. In certain cases pictures were not fully recoverable, but they helped demonstrate the interests and activity being performed during the browsing session.
4. What evidence was recoverable during post-mortem analysis in relation to Browzar?
   Using X-Ways and IEF as primary tools, several additional artifacts were recovered. Files and folders left behind by Browzar were recovered using manual techniques from within X-Ways. The WebCacheV01.dat and Pagefile.sys files proved to contain a wealth of information - including evidence of Browzar keyword searches, websites visited, and pictures.

## 5     Conclusions and future work

    In this paper, we examine Browzar, which claims to be a browser specifically designed for privacy, comparing results to the built in private browsing modes in Google Chrome and Mozilla Firefox Internet browsers. This chapter also provided elaborate and detailed answers relating to the forensically recoverable information for each web browser tested. The results of this research are useful to, and may be referenced by, forensic experts involved in investigations concerning web activity on both desktop and mobile platforms [17, 18, 19, 20]. Each question posed in the problem statement was addressed and the underlying data presented in subsequent sections should help not only investigators looking for a better general understanding of web browser forensics, but also those seeking advanced techniques and methods for recovering, parsing and analysing web browser specific data. Some topics for further scientific and practical research are coming up. First of all, investigators can use forensic method proposed in this chapter to examine other privacy Internet browsers such as Epic Privacy Browser (https://www.epicbrowser.com/). Next, this research also assists the studying of website fingerprinting [21, 22]. Besides, we are looking at using clustering methods and tree-based approach [23, 24, 25] to group correlated artifacts from Browzar to deeply analyze the evidence. Moreover, experimental results described in this paper can assist the researchers who are studying for new methods of preserving the privacy in the next generation of web browser and mobile apps [26].

## References

1. Junghoon Oh a,*, Seungbong Lee b , Sangjin Lee Advanced evidence collection and analysis of web browser activity, The Digital Forensic Research Conference, DFRWS 2011 USA, New Orleans, LA (Aug 1- 3), 2011
2. Larkin, E. (December 01, 2008). Privacy Watch. Pc World, 26, 12, 50
3. PC World. "How Private—or Secure—Is so-Called Private Browsing?" [Online]. Available http://www.pcworld.com/article/152966/private_browsing.html. [Accessed: 30-December-2016].
4. Internet Live Stats - http://www.internetworldstats.com/emarketing.htm
5. Internet Live Stats - http://www.internetlivestats.com/google-search-statistics/
6. Interpol. "Cybercrime". [Online]. Available http://www.interpol.int/Crime-areas/Cybercrime/Cybercrime. [Accessed: 30-December-2016].
7. Europol. "Europol Identifies 3600 Organised Crime Groups Active in the EU". [Online]. Available https://www.europol.europa.eu/content/europol-identifies-3600-organised-crime-groups-active-eu-europol-report-warns-new-breed-crim. [Accessed: 30-December-2016].
8. Piriform, CCleaner, https://www.piriform.com/ccleaner [Accessed: 30-December-2016].
9. http://www.bigplanetusa.com/library/bp/pdf/bpis_understanding_security.pdf
10. Oh, J., Lee, S., & Lee, S. (2011). Advanced evidence collection and analysis of web browser activity. Digital Investigation 8, s62-70
11. www.apple.com/safari
12. Gaurav , Aggarwal, Ellie, Bursztein, Collin Jackson, Dan Boneh - An Analysis of Private Browsing Modes in Modern Browsers, USENIX Security'10 Proceedings of the 19th USENIX conference on Security, Washington DC, August, 2010
13. Ashley Hedberg, The Privacy of Private Browsing, Technical Report, http://www.cs.tufts.edu/comp/116/archive/fall2013/ahedberg.pdf
14. Sandeep Kumar Khanikekar. (2010). Web Forensics. Graduate Thesis, A&M University, Texas
15. https://www.epicbrowser.com/
16. http://www.browzar.com/
17. Sgaras C., Kechadi MT., Le-Khac NA. (2015) Forensics Acquisition and Analysis of Instant Messaging and VoIP Applications. In: Garain U., Shafait F. (eds) Computational Forensics. Lecture Notes in Computer Science, Vol 8915. Springer, Cham
18. Faheem M., Kechadi MT., Le-Khac NA. (2016), Toward a new mobile cloud forensic framework 6th IEEE International Conference on Innovative Computing Technology, Ireland, 2016
19. Faheem M., Kechadi MT., Le-Khac NA. (2015) The State of the Art Forensic Techniques in Mobile Cloud Environment: A Survey, Challenges and Current Trends, International Journal of Digital Crime and Forensics (IJDCF), Vol.7 (2) pp.1-19
20. Hitchcock B., Le-Khac NA., Scanlon M. (2016) Tiered Forensic Methodology Model for Digital Field Triage by Non-Digital Evidence Specialists, Digital Investigation Vol.16(29), 2016, DOI: https://doi.org/10.1016/j.diin.2016.01.010
21. Acar G., et al. (2014). The Web never forgets: Persistent tracking mechanisms in the wild. In Proceedings of CCS 2014, Nov. 2014
22. Juarez M. et al. (2016) Toward an Efficient Website Fingerprinting Defense. In: Askoxylakis I., Ioannidis S., Katsikas S., Meadows C. (eds) Computer Security – ESORICS 2016. ESORICS 2016. Lecture Notes in Computer Science, Vol.9878. Springer, Cham

23. Le Khac NA. et al. (2010) A Cluster-based Data Reduction for Very Large Spatio-Temporal Datasets International Conference on Advanced Data Mining and Applications, China, Nov, 2010
24. Le-Khac NA. et al. (2009) An Efficient Search Tool for an Anti-Money Laundering Application of an Multi-National Bank's Dataset. In: CESRA Press eds. 2009 International Conference on Information and Knowledge Engineering (IKE'09) Las Vegas, USA, July 2009
25. Aouad, L-M. et al. (2009) 'Grid-based Approaches for Distributed Data Mining Applications'. Journal of Algorithms & Computational Technology, Vol 3(4), 2009
26. Voorst, RV.; Kechadi, T.; Le-Khac, NA. (2015) 'Forensics Acquisition of Imvu: A Case Study'. Journal of Association of Digital Forensics, Security and Law, Vol.10 (4), 2015